\documentclass{optica-article}

\journal{opticajournal}

\articletype{Research Article}
\usepackage{xcolor}
\usepackage{lineno}
\usepackage{float}


\AtBeginDocument{\captionsetup{width=\textwidth}}

\begin{document}

\title{Intensity-Domain Demultiplexing of Orbital Angular Momentum Channels for Optical Communication} 

\author{Junsu Kim\authormark{1}, Jiwon Jung\authormark{2}, Hyunchae Chun\authormark{2*}, and Seung Ryong Park\authormark{1*}}

\address{\authormark{1,1*}Department of Physics, Incheon National University (INU), Incheon, South Korea}

\address{\authormark{2,2*}Department of Information and Telecommunication Engineering, Incheon National University (INU), Incheon, South Korea}

\email{\authormark{1*}abepark@inu.ac.kr, \authormark{2*}hyunchae.chun@inu.ac.kr}

\begin{abstract*}
Orbital angular momentum (OAM) multiplexing has been limited in practice by the complexity and optical loss of the elements that separate the channels at the receiver side. Here we adopt the azimuthal harmonics of OAM petal patterns as the demultiplexing basis, which makes the receiver the optical counterpart of the circular antenna arrays used in radio OAM links: the petal pattern of topological charge $\ell$ carries a single azimuthal harmonic of order $2\ell$, so one Fourier transform of the azimuthal intensity profile returns every channel, without gratings or mode sorters. The petals are formed by projecting a vector vortex beam from a single q-plate onto a linear polarization. A two-channel system at 515 nm with a 16-position circular photoreceiver array reaches 5.2 Gb/s below the forward error correction threshold, 2.6 Gb/s per channel from a 1.4 GHz photoreceiver.
\end{abstract*}

\section{Introduction}

Orbital angular momentum (OAM) offers communication links an additional dimension, and its promise of more channels and higher data rates on a single beam has made OAM multiplexing an active research topic \cite{Willner15,Willner21,Trichili19,Gibson04}. As a degree of freedom of light, OAM is most naturally described in the Laguerre-Gaussian (LG) basis, whose helical wavefronts have earned these modes the names optical vortices and twisted photons \cite{Allen92,Yao11,Padgett17,Stoyanov23,MolinaTerriza07,Erhard17,Baumann09,Vickers08,Paufler19}. What characterizes OAM is the azimuthal phase factor $e^{i\ell\phi}$, where the integer $\ell$, the topological charge, sets how many times the phase winds around the beam axis and fixes the OAM carried per photon \cite{Leach02,Mair01,Kotlyar17,Zheng17}. Because modes with different $\ell$ are mutually orthogonal, each can carry an independent data stream. Realizing this capacity depends on how well the channels are combined at the transmitter and separated again at the receiver \cite{Wang12,Huang14,Bozinovic13,Yan14,Shi22}, and it is the separation, demultiplexing, that limits the added dimension in practice.

Most wireless communication runs on radio and microwave carriers, where OAM multiplexing has also been introduced and demultiplexing poses little difficulty, because the receiver measures the phase of the field directly. A uniform circular antenna array samples the complex field at equally spaced points on a ring, and a discrete Fourier transform of these samples separates the modes, since the phase factors $e^{i\ell\phi}$ of different $\ell$ are orthogonal on the ring \cite{Chen20}. The separation is a spectral analysis, the same operation that resolves a radio signal into its frequency components, carried out along the azimuth instead of in time, and such array receivers are closely related to the multiple-input multiple-output (MIMO) systems that are standard in radio communication. Optical OAM links are pursued just as actively, but at optical frequencies the phase of the field cannot be measured directly. A detector records intensity, $|E|^{2}$, which is quadratic in the field, and under that map the modes no longer add as orthogonal components. The azimuthal intensity profile of a single mode is uniform and retains no sign of $\ell$, so once several channels share a beam, their contributions cannot be told apart from the recorded profile.

Optical demultiplexing has therefore relied on an element that sorts the modes in the spatial domain before detection, such as diffraction gratings, metasurfaces, refractive modulators or mode sorters, or, more recently, on neural networks acting on the recorded image \cite{Feng22,Lei15,Leith62,Heckenberg92,Arlt98,Stoyanov15,Topuzoski19,Ren19,Yu24,Longman17,Berkhout10,Mirhosseini13,Fontaine19,Fang22,Ye24,Wang21}. The most widely used of these, holographic gratings displayed on spatial light modulators \cite{Heckenberg92} and refractive mode sorters \cite{Berkhout10,Mirhosseini13}, illustrate the cost: the former loses power to diffraction and requires an active device, and the latter requires custom optics aligned with high precision. In general, each approach adds an optical or a computational stage that increases the cost and complexity of the receiver side. Atmospheric turbulence adds a further difficulty by scattering power between modes, and correcting this inter-modal crosstalk is an active field in its own right \cite{Paterson05,Ren13,Malik12,Manavalan25,Chen25,Liu26}. The receiver side complexity and loss weigh most heavily on free-space optical (FSO) links between the ground and satellites and between satellites, where optical communication is of particular value, loss over the long path is critical, and the simplicity of the receiver side is itself a key bottleneck \cite{Kaushal17,Ahmad26}. For these reasons, OAM multiplexing has yet to reach practical optical links.

We address these problems with a demultiplexing method that receives the beam directly on a ring-shaped detector array and separates the channels with a single Fourier transform, the optical counterpart of the radio receiver. Each channel is carried by a petal pattern with $2\ell$ lobes around the beam axis, the intensity of the OAM pair state $|\psi_{\text{pair}}^\pm\rangle=(|+\ell\rangle\pm|-\ell\rangle)/\sqrt{2}$, where $|\ell\rangle$ is the OAM eigenstate. This pattern contains exactly one azimuthal harmonic, of order $2\ell$, so different $\ell$ occupy different harmonics, and these harmonics are orthogonal in the intensity domain in the same way that the modes are orthogonal in the field domain. The petal pattern has a direct analogue in the real spherical harmonics of atomic orbitals: just as $Y_l^{m}\pm Y_l^{-m}$ turns the uniform density of $Y_l^{m}\propto e^{im\phi}$ into $\cos^2(m\phi)$ or $\sin^2(m\phi)$ lobes, with $\ell$ playing the role of $m$, the orthogonality of the wavefunctions becomes visible in the lobe pattern. By this orthogonality, a single Fourier transform of the azimuthal intensity profile returns the amplitude of every channel at once. The intensity must be sampled at several points on the ring, but no sorting element enters the beam path, and the cost and loss of such an element are avoided. We introduced this concept in our previous work \cite{Kim26}, where it was verified only numerically. An optical pair mode is normally formed either by superposing separately generated $+\ell$ and $-\ell$ beams in an interferometer, where the relative phase between the two arms sets the petal orientation, so the orientation drifts with it, or by displaying the pair-state hologram on a spatial light modulator, which costs diffraction efficiency and adds an active element. 

Here, we adopt a different way to generate petal patterns, by a linear polarization projection of a vector vortex beam (VVB) generated by a single q-plate. The two components travel on orthogonal polarizations of the same beam, so their relative phase is fixed by the polarization state, and the petal pattern appears behind a polarizer, with its orientation set by the polarizer angle. The q-plate ($2q=\ell$) fixes $\ell$ for each channel, which is the natural condition of a link. We demonstrate the principle experimentally in a two-channel system with $\ell=1$ ($q=1/2$) and $\ell=2$ ($q=1$) at 515 nm, received on a 16-position circular photoreceiver array, reaching 5.2 Gb/s below the forward error correction threshold.

Petal patterns have been used before in communication and security. They have been recognized by correlating the received intensity with a library of reference patterns \cite{Rumman26}, classified by a trained neural network in camera communication, with the radial and azimuthal orders together encoding a 5-bit symbol \cite{Chuo25}, and used as keys of a multiplexed hologram, in which the petal order and orientation unlock one encrypted image at a time \cite{Lin25}. In each case a beam path carries a single petal pattern at a time, and the pattern is a symbol to be recognized or a key to be matched; where several patterns are sent together, they occupy separate positions. In this paper, petal patterns of different $\ell$ share the same beam, each carrying its own data stream as an independent channel, and are separated by their azimuthal harmonics: one Fourier transform of the photoreceiver samples returns the amplitudes of all channels at once, without a reference library or a trained model. The details are given in the next section.

\section{Theory \& Experiments}
OAM petal patterns can be generated by projection of VVB \cite{Marrucci06,Yao25,Yu24b,Milione15,Gao25}. The method which we adopt in this work, uses VVB, which can be represented as
\begin{equation}
    |\Psi_{\text{VVB}}\rangle=\frac{1}{\sqrt{2}}(|L\rangle\otimes|+\ell\rangle+|R\rangle\otimes|-\ell\rangle)=|H\rangle\otimes\left(\frac{|+\ell\rangle+|-\ell\rangle}{2} \right)+|V\rangle\otimes\left(\frac{|+\ell\rangle-|-\ell\rangle}{2i} \right)
\end{equation}
, where $|L\rangle=(|H\rangle-i|V\rangle)/\sqrt2$ and $|R\rangle=(|H\rangle+i|V\rangle)/\sqrt2$ are left and right circular polarization states, and $|H\rangle$ and $|V\rangle$ are horizontal and vertical linear polarization states. Eq. (1) represents a state of maximal spin-orbit coupling, in which the polarization and OAM degrees of freedom cannot be factorized. The tensor product right-side term is a form of pair mode state as linear combination of opposite sign. The beam itself carries no azimuthal intensity pattern. A linear polarizer converts it into a petal pattern whose orientation follows the polarizer angle. Projecting the VVB state onto an azimuthal phase state $|\phi\rangle$ and polarization state $|P\rangle$ gives
\begin{equation}
\begin{aligned}
\left(\langle P|\otimes\langle \phi|\right)|\Psi_{\text{VVB}}\rangle
&=\frac{1}{2\sqrt{2\pi}}\left[\langle P|H\rangle\left(e^{i\ell\phi}+e^{-i\ell\phi}\right)+\langle P|V\rangle\left(\frac{e^{i\ell\phi}-e^{-i\ell\phi}}{i}\right)\right]\\&=\frac{1}{\sqrt{2\pi}}\left[\langle P|H\rangle\cos(\ell\phi)+\langle P|V\rangle\sin(\ell\phi)\right]
\end{aligned}
\end{equation}
, where $\langle\phi|\pm\ell\rangle\equiv\left(1/\sqrt{2\pi}\right)e^{\pm i\ell\phi}$, and $\langle P|H\rangle$ and $\langle P|V\rangle$ are the inner products of the analyzed polarization state $|P\rangle$ with the basis states $|H\rangle$ and $|V\rangle$. For a linear polarizer oriented at an angle $\alpha$ from the horizontal axis,
\begin{equation}
|P\rangle=\cos(\alpha)|H\rangle+\sin(\alpha)|V\rangle
\end{equation}
We can rewrite the projection of the VVB state onto the azimuthal phase state and polarization state as
\begin{equation}
\left(\langle P|\otimes\langle \phi|\right)|\Psi_{\text{VVB}}\rangle=\frac{1}{\sqrt{2\pi}}\left[\cos(\alpha)\cos(\ell\phi)+\sin(\alpha)\sin(\ell\phi)\right]=\frac{1}{\sqrt{2\pi}}\cos(\ell\phi-\alpha)
\end{equation}
The intensity profile of the projected state is given by
\begin{equation}
I^{(P)}_{\text{VVB}}\equiv A_{\ell} \cos^2(\ell\phi-\alpha)
=\frac{A_\ell}2[\cos(2\ell\phi-2\alpha)+1]
\end{equation}
, where $A_\ell$ is an amplitude factor. Eq. (5) is the basis of the demultiplexing method proposed in this work. The projected intensity carries a single azimuthal harmonic, of order $2\ell$ and phase $2\alpha$. The topological charge is thereby mapped onto a Fourier index of the azimuthal coordinate, so that beams of different $\ell$ occupy mutually orthogonal harmonics and can be separated by a single Fourier transform over $\phi$. The constant term is common to every channel and carries no information about $\ell$. It can be removed by subtracting the azimuthal mean $I_{\text{mean}}$, as
\begin{equation}
\tilde{I}^{(P)}_{\text{VVB}}(\phi)\equiv I^{(P)}_{\text{VVB}}(\phi)-I_{\text{mean}}=\frac{A_\ell}{2}\cos(2\ell\phi-2\alpha)
\end{equation}
The Fourier transform of the mean-subtracted intensity profile is given by
\begin{equation}
H_p\equiv\frac{1}{\pi}\int_{0}^{2\pi}\tilde{I}^{(P)}_{\text{VVB}}(\phi)\,e^{ip\phi}\,d\phi=\frac{A_\ell}{2}\,e^{i2\alpha}\,\delta_{p,2\ell}
\end{equation}
, so that a beam of topological charge $\ell$ occupies a single petal harmonic index as $p=2\ell$, with amplitude $A_\ell/2$ and phase $2\alpha$. The orthogonality of the OAM basis is thus restored in the intensity domain. It is not necessary to align $\alpha$ for each OAM mode, since the phase term is irrelevant to the $p$ harmonic index. The method can be extended to any number of OAM channels, by multiplexing the petal patterns of different $\ell$ into a single beam and taking the Fourier transform of the azimuthal intensity profile. Each channel is thereby demultiplexed into a single harmonic index $p=2\ell$, with amplitude $A_\ell/2$.

\begin{figure}[!t]
    \centering
    \includegraphics[width=\linewidth]{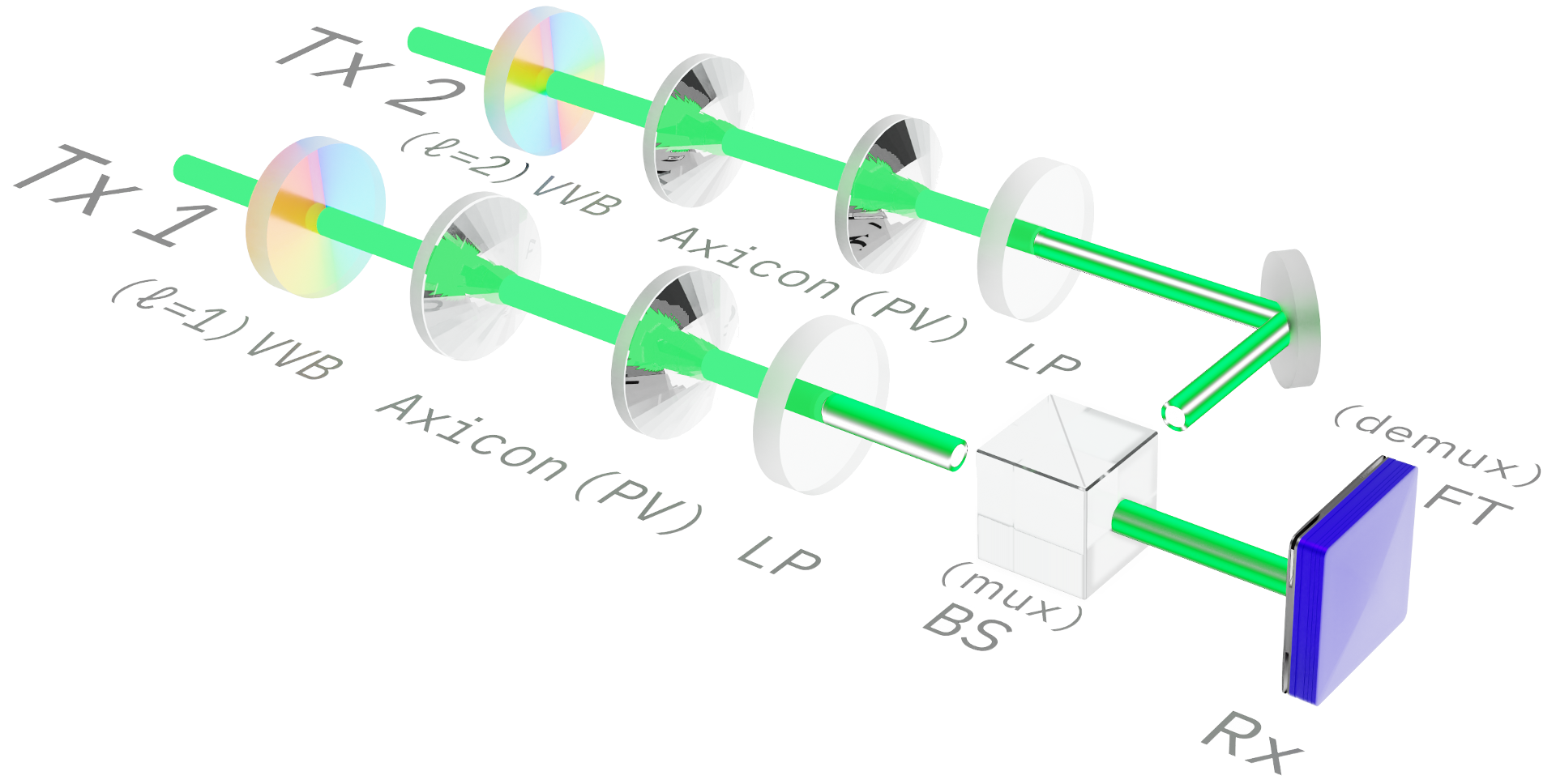}
    \caption{The experimental setup for OAM two-channel Tx 1 (channel 1) and Tx 2 (channel 2) communication system. The vector vortex beam (VVB) is generated by a q-plate. Two axicons are used to generate collimated Perfect Vortex (PV) beam. Linear polarizer projects the VVB into a petal pattern for each mode. Each petal pattern $\ell=1$ and $\ell=2$ are multiplexed. The mux beam is detected by a camera and 16 points circular photoreceiver array. The mux petal pattern is demultiplexed by its own harmonics analytically.}
    \label{fig:experimental_setup}
\end{figure}

This harmonic property can be stated in terms of the rotational symmetry of the projected field \cite{Sakurai20}. Let $\hat{R}(\theta)$ rotate the transverse plane about the propagation axis, $\hat{R}(\theta)|\phi\rangle=|\phi+\theta\rangle$, so that
\begin{equation}
\hat{R}(\theta)\,|\pm\ell\rangle=e^{\mp i\ell\theta}\,|\pm\ell\rangle
\end{equation}
Writing $|\pi_\ell(\alpha)\rangle\equiv\langle P|\Psi_{\text{VVB}}\rangle=\frac12(e^{-i\alpha}|+\ell\rangle+e^{+i\alpha}|-\ell\rangle)$ for the projected field of Eq. (4), both of its $|{\pm}\ell\rangle$ components acquire a factor of $-1$ at $\theta=\pi/\ell$, so that
\begin{equation}
\hat{R}(\pi/\ell)\,|\pi_\ell(\alpha)\rangle=-\,|\pi_\ell(\alpha)\rangle
\end{equation}
The projected field is therefore antisymmetric under a rotation by $\pi/\ell$ and returns to itself only after $2\pi/\ell$, whereas the intensity, being square of the field, is already invariant under $\pi/\ell$. The projected intensity thus carries the cyclic symmetry $C_{2\ell}$ while the field carries only $C_\ell$, and this doubling of the rotational order is the origin of the factor two in $p=2\ell$. The receiver measures intensity rather than field, the accessible symmetry is $C_{p}$, and each channel is addressed through the single harmonic of that order without any access to the optical phase.

There is an additional advantage from its own property. The method reads only even harmonics, $p=2\ell$, and this parity gives it a built-in immunity to a whole class of disturbances. The parity of the spurious harmonics provides a useful diagnostic of the dominant imperfection. A one-fold perturbation, such as a lateral displacement of the beam with respect to the photoreceiver array, an illumination tilt, or a wedge-like intensity gradient across the aperture, has odd-order harmonics ($p=1,3,\dots$) as its Fourier components. These are orthogonal to even harmonics of $p=2$ and $p=4$ and never enter a channel. What the parity does not reject are even order distortions.

Petal pattern is generated by a polarization state projection of VVB, but it does not mean that the polarization state is used for channel dimension. The polarization state projection is only for generating the petal pattern. Fig. 1's linear polarizer is separated to show the petal pattern generation process. The linear polarizer can be placed right before detection. The polarizer projects the vector vortex beam onto a single polarization state, so the beam leaving it is uniformly polarized and carries the channel information entirely in its azimuthal intensity pattern. Polarization stays as a free degree of freedom once the petal pattern has been formed. For instance, two petal beams multiplexed in this way can be given orthogonal polarizations, and at the receiver a polarizing beam splitter separates them again into two ways. Each way then demultiplexes its own full set of OAM channels by the azimuthal Fourier transform, so polarization and OAM multiply rather than compete, and the channel count becomes twice that of a single array. This paper mainly focuses on the capability of the independent OAM azimuthal harmonic demultiplexing method. The cost of this simplification is on the detection side. The receiver must be built as a circular array of photoreceivers, each with the full signal bandwidth, instead of one photodetector per mode. The array is shared by all channels, and its size is set by the highest topological charge in use rather than by the number of channels, since $N_{\rm d}\ge4\ell_{\max}$. Separating the OAM channels requires no additional optical components, because the demultiplexing is carried out entirely as a Fourier transform of the signals after detection.

We demonstrate this method experimentally as shown in Fig. 1. We adopt channel 1 (Tx 1) as $\ell=1$ state and channel 2 (Tx 2) as $\ell=2$ state, and multiplex them into a single beam line. Two 515 nm laser diodes are used, which are modulated by an arbitrary waveform generator (Tektronix AWG70002A Opt. 216, 16 GS/s). The VVB is generated by a q-plate (Thorlabs Zero-Order Vortex Half-Wave Retarders). Two axicons are used to generate collimated Perfect Vortex (PV) beam for ideal petal pattern generation \cite{Vaity15,Liu17,PereiroGarca23,Topuzoski09,Karahroudi17}. Axicon's physical angle is $5^\circ$, and manufactured from UV Fused Silica (Thorlabs AX255M-A). Beam diameter is 8 mm.  Linear polarizer projects the VVB into a petal pattern for each mode what we mentioned before. Two linear polarizers project each channel into same polarization state in our setup. The link distance is about 0.1 m for 16 photoreceiver array and 0.5 m for CMOS camera. Finally, the multiplexed petal pattern is detected by a CMOS camera and 16 points circular photoreceiver array. We use (Femto HSA-X-S-1G4-SI-FST) photoreceiver which has 400 $\mu$m active area, and 10 kHz to 1.4 GHz bandwidth. The photoreceiver output is recorded with a 4 GHz oscilloscope (Tektronix DPO70404C) at 16 GS/s. Detected signal is processed by our method.

\begin{figure}[!tb]
    \centering
    \includegraphics[width=\linewidth]{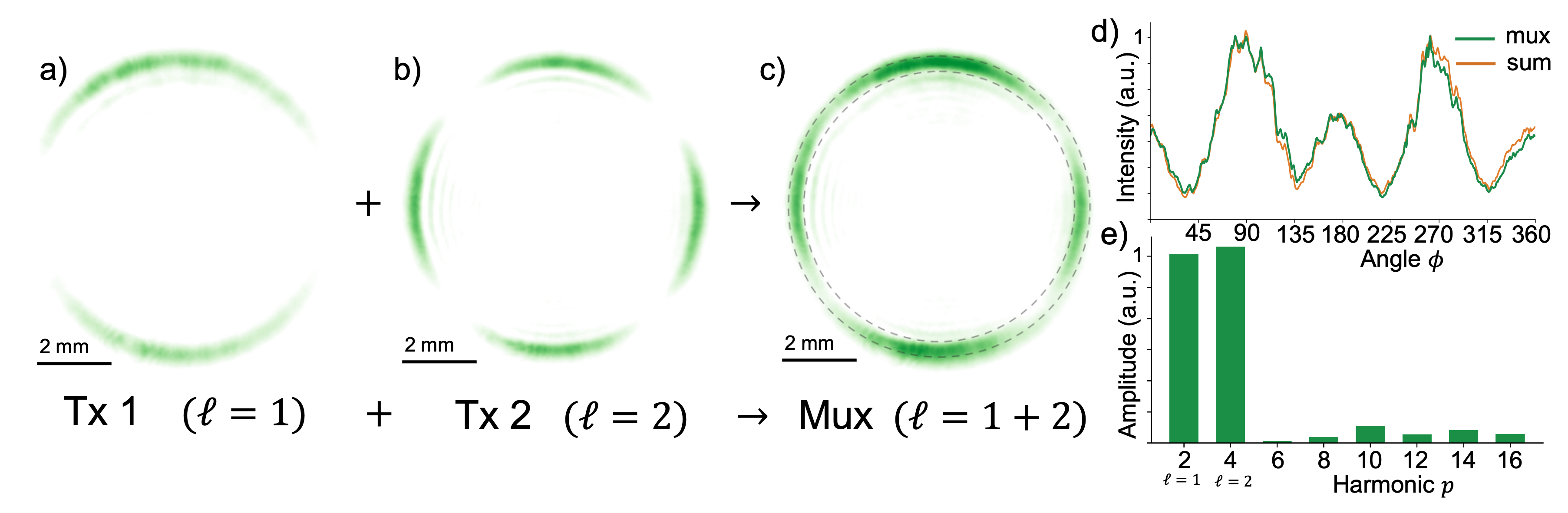}
    \caption{CMOS camera detected images. (a) Tx 1 ($\ell=1$) petal pattern intensity profile. (b) Tx 2 ($\ell=2$) petal pattern intensity profile. (c) the multiplexed intensity profile of Tx 1 and Tx 2. (d) the azimuthal intensity profile of (c) at its effective radius, taken in polar coordinates over the annulus marked by the dashed circles. The green solid line is multiplexed profile and orange solid line is the intensity sum of $\ell=1$ and $\ell=2$ profile. (e) the demultiplexed amplitude which is obtained by subtracting the azimuthal mean and taking the Fourier transform of (d). Its amplitude value for each harmonic ($p=2$ or $p=4$) becomes each channel's amplitude.}
    \label{fig:petal_patterns}
\end{figure}

Fig. 2 shows the CMOS camera (Thorlabs CS165MU) detected images of petal patterns of $\ell=1$ and $\ell=2$. They are grayscale captures, which are displayed in green false color, matching the 515 nm source. The multiplexed intensity profile is shown in Fig. 2(c), and its azimuthal intensity profile is shown in Fig. 2(d). The green solid line is multiplexed profile and orange solid line is the intensity sum of $\ell=1$ and $\ell=2$ profile. Intensity differences between them can be negligible. We can demultiplex the two channels by taking the Fourier transform of the azimuthal profile, as shown in Fig. 2(e). The amplitude value of each harmonic ($p=2$ or $p=4$) becomes each channel's amplitude. Of the total harmonic power in the multiplexed profile, 94 \% lies in the two harmonics $p=2$ and $p=4$, and 4.7 \% in the odd harmonics $p=1$ and $p=3$, and 1.3 \% in all orders above four. The channel information is therefore concentrated in two harmonic indices.

Usual FSO communication is based on non coherence condition. However there can be coherence effect between two channels, the summation has to be considered in field domain. In our two channels case $p=2$ and $p=4$, the coherence correlation term is only positioned on odd terms 1 and 3 harmonics. So it does not disturb communication process. In general two channel scenario, only $\ell'=3\ell$ cases coherence terms disturb original harmonics. This paper's system is not affected even if there is mutual coherence. The general equation is derived in Appendix A.

\section{Results \& Discussion}
The circular photoreceiver array is composed of 16 positions at equal azimuthal intervals (22.5$^{\circ}$) on a circle of diameter 8 mm (Fig. 3 (a) inset). The ring is sampled by a single photoreceiver moved through the sixteen positions in turn, recording the full transmitted sequence at each. Because the same bit pattern is sent every time, the sixteen recordings are synchronized to it on one time base, and the demultiplexing is carried out on the aligned set in post-processing. We represent three-dimensional data stream by plotting the azimuthal intensity modulation profile over time. For each time instance, Fourier transform is taken along the azimuthal axis. Each OAM mode channel is separated as we mentioned before. Fig. 3(a) shows the azimuthal intensity modulation profile of the two-channel multiplexed data stream over 10 bit periods at 1 Gb/s per each channel. It is reconstructed from the $p=2$ and $p=4$ harmonic components for showing clear modulation patterns. Fig. 3(b) shows the demultiplexed amplitude data stream of each channel's azimuthal harmonic. Dashed lines are the binary tx reference data stream, and solid lines are the rx data stream. We design binary data stream as pseudo random binary sequence (PRBS) for each channel as channel 1 to PRBS-15 and channel 2 to PRBS-23. Using the demultiplexed data stream, we collect the number of error bits and calculate the Bit Error Rate (BER) of each channel. Repeat this process for different data rates, and we find forward error correction (FEC)-limited maximum data rate point between the total BER and the FEC threshold \cite{Milione15,Zuo26,Hei23}. Fig. 3(c) shows the BER of channel 1 and channel 2. Each channel has $3\times 10^{5}$ bits. Fig. 3(d) shows the total BER which is calculated by the sum of channels' error bits with $6\times 10^{5}$ bits. Data rate 5.2 Gb/s is the FEC-limited maximum data rate between the total BER and the FEC threshold in this system. The inset shows the eye diagram of each channel at 4.0 Gb/s which is onset point. 

\begin{figure}[t]
    \centering
    \includegraphics[width=\linewidth]{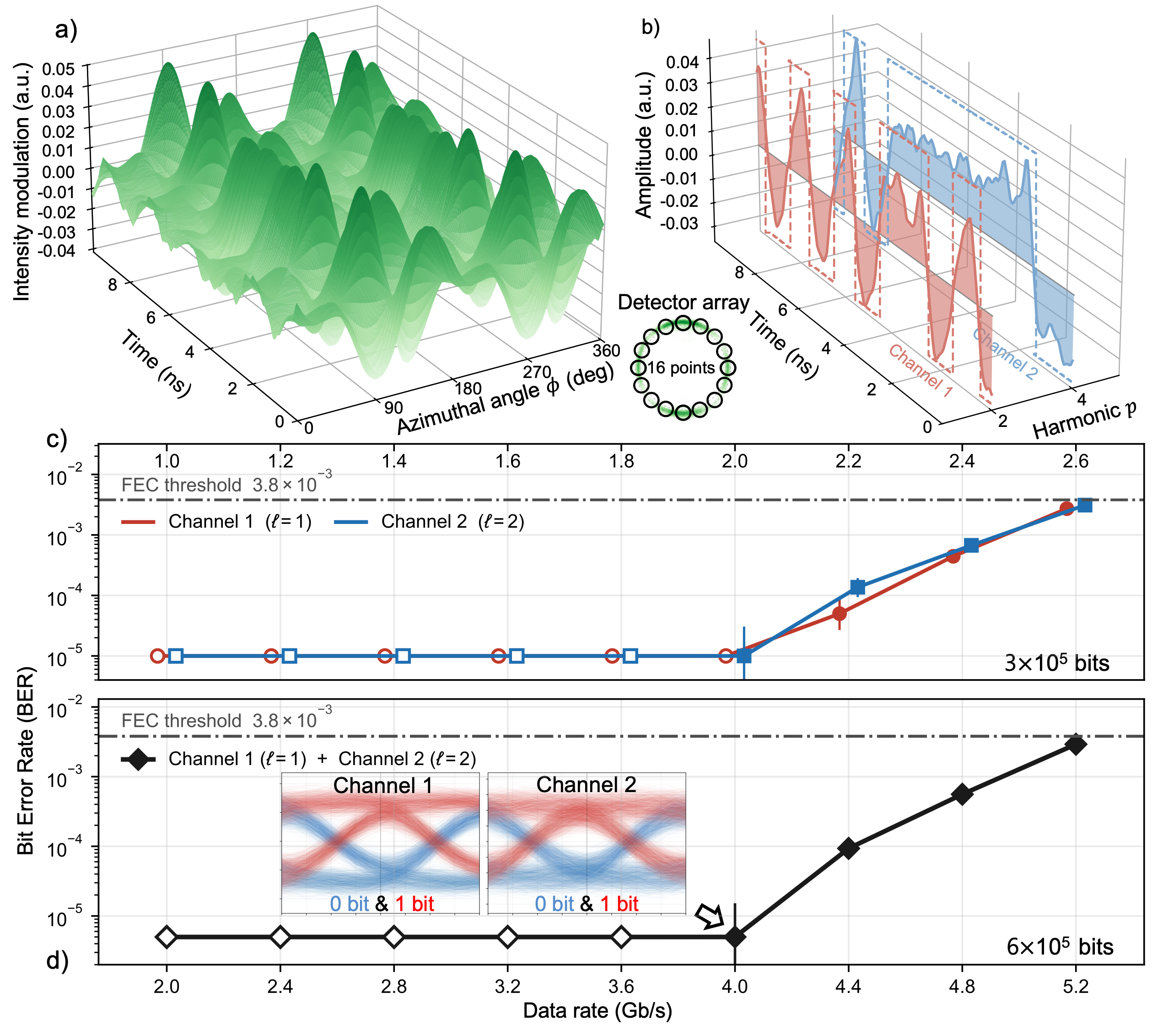}
    \caption{The detected signal processed by our method. (a) the azimuthal intensity modulation profile of the two-channel multiplexed data stream over 10 bit periods at 1 Gb/s per each channel, reconstructed from the $p=2$ and $p=4$ harmonic components of the 16 points circular photoreceiver array (inset). (b) the demultiplexed amplitude data stream of each channel's azimuthal harmonic. Dashed lines are the tx reference data stream, and solid lines are the rx data stream. (c) Tx 1 becomes channel 1 and Tx 2 becomes channel 2. The BER of channel 1 and channel 2 are calculated by the demultiplexed amplitude of each channel. Each channel has $3\times 10^{5}$ bits. (d) Total BER is calculated by the sum of channel 1 and channel 2's error bits with $6\times 10^{5}$ bits. Open markers denote rates at which no errors were detected and are plotted at $3/N$, the 95\% upper bound for zero errors in $N$ bits ($1\times10^{-5}$ in (c), $5\times10^{-6}$ in (d)); filled markers are measured BERs, and vertical bars give the 95\% Poisson confidence interval of the error count. 5.2 Gb/s is the FEC-limited maximum data rate point between the total BER and the FEC threshold in this system. The inset shows the eye diagram of each channel at 4.0 Gb/s.}
    \label{fig:3D_profile}
\end{figure}

For example at the FEC-limited point of 2.6 Gb/s per channel, there are 816 errors / $3\times 10^{5}$ bits on channel 1 and 934 errors / $3\times 10^{5}$ bits on channel 2. The corresponding signal to noise ratio (SNR), obtained from the measured BER through $\mathrm{BER}=\tfrac12\mathrm{erfc}(Q/\sqrt2)$ and $\mathrm{SNR}=20\log_{10}Q$, is 8.88 dB on channel 1 and 8.74 dB on channel 2. Whether these values are set by crosstalk between the channels or by the photoreceiver is examined next.

We quantify the crosstalk between the channels. Its definitions are in Appendix B. Over the nine data rates, $\mathrm{XT}_1$ ranges from $-24.3$ to $-17.6$ dB and $\mathrm{XT}_2$ from $-8.8$ to $-7.3$ dB. The former is comparable to or lower than the $-10$ to $-20$ dB reported for two- to four-mode OAM communication in free space \cite{Zuo26,Wang20,Milione15,Li17} and under water~\cite{Hei23}. The latter is larger, and can be removed after detection by subtracting the contribution of channel 1,
\begin{equation}
s_2'=s_2-M_{21}\hat{b}_1
\end{equation}
, where $\hat{b}_1$ is the channel 1 decision, which lowers $\mathrm{XT}_2$ below $-27$ dB. All BER values in Fig. 3 and the 5.2 Gb/s rate are obtained without this subtraction. The two SNRs are therefore close for different reasons. The crosstalk into channel 1 costs less than 0.2 dB of SNR, so 8.88 dB is its intrinsic SNR. Channel 2 has a lower noise and an intrinsic SNR of 10.15 dB, obtained from the 193 errors that remain after the subtraction, but the crosstalk from channel 1 costs 1.41 dB and brings it to 8.74 dB.

Both differences have a definite origin. The crosstalk is asymmetric because of the symmetry of the patterns. The $C_2$-symmetric pattern of channel 1 is allowed a $p=4$ component, and any departure of its petals from a pure $\cos^2$ shape produces one, which the receiver reads as channel 2, whereas the $C_4$-symmetric pattern of channel 2 is forbidden a $p=2$ component, so the reverse is suppressed. In general, the pattern of channel $\ell'$ can place power on channel $\ell$ only when $\ell'$ divides $\ell$, apart from aliasing of harmonics above the sampling limit. The noise is asymmetric because of the transmitters, not the patterns. The Fourier transform over the 16 positions places the same photoreceiver noise in every harmonic, so the number of petals does not enter, and the difference comes from the slower edge of channel 1 ($p=2$), whose intersymbol interference grows faster with the data rate.

Channel 1, which has no crosstalk to remove, still produces 816 errors at 2.6 Gb/s, so the FEC-limited point is set by the photoreceiver rather than by crosstalk. The photoreceiver's responsivity peaks at 0.51 A/W at 760 nm and is about 0.25 A/W at 515 nm. The 2.6 Gb/s data rate is 1.86 times the 1.4 GHz bandwidth, and over the range from 1.0 to 2.6 Gb/s the modulation depth of channel 1 falls by 32 \% while its noise rises from 3.57 mV to 8.35 mV, and channel 2 falls by 29 \% while its noise rises from 5.24 mV to 6.60 mV. The FEC-limited point can therefore be raised with a photoreceiver of higher bandwidth and responsivity together with a signal amplifier.

\begin{figure}[!t]
    \centering
    \includegraphics[width=\linewidth]{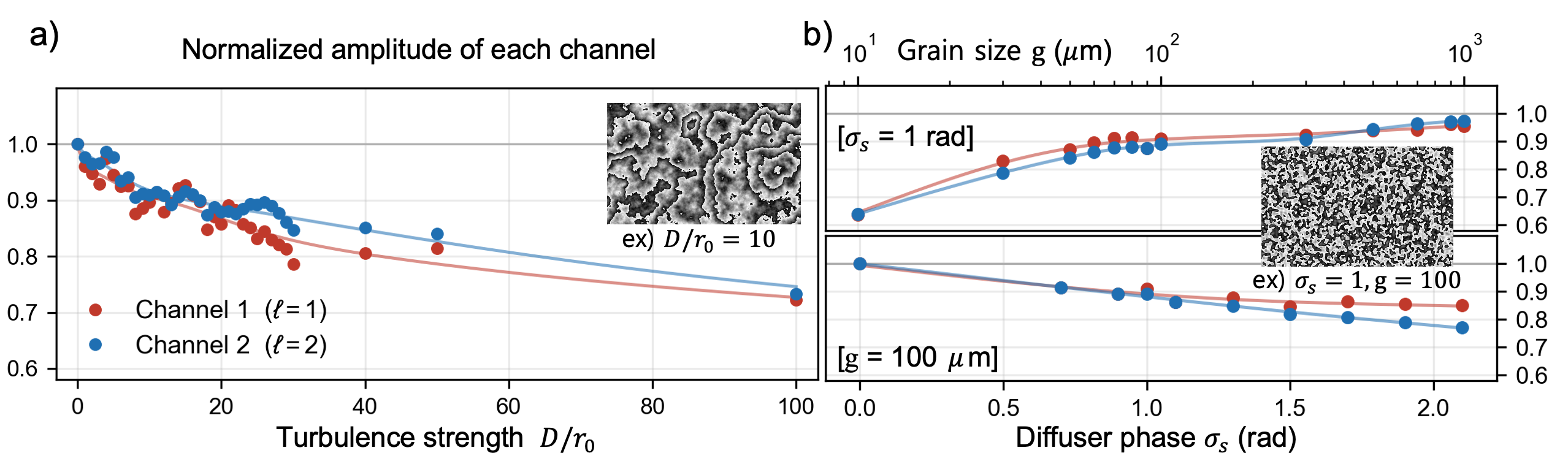}
    \caption{Degraded amplitude scenarios. (a) normalized amplitude of each channel about turbulence strength $D/r_0$. Inset is an example at parameter $D/r_0=10$. (b) normalized amplitude of each channel about speckle strength as diffuser phase $\sigma_s$ at bottom and grain size $\rm g$ at top. Inset is an example at parameter $\sigma_s=1$ rad and $\rm g=100\ \mu$m. Red dot, solid line is channel 1, and blue dot, solid line is channel 2.}
    \label{turblence}
\end{figure}

Additionally, we demonstrate the effect of turbulence and speckle on the demultiplexing method. We use a spatial light modulator (SLM, Thorlabs EXULUS-HD2) to generate holographic of turbulence phase screen and speckle. Its resolution is 1920 $\times$ 1200, pixel pitch is 8 $\mu$m, and panel active area is 15.42 mm $\times$ 9.66 mm and beam diameter is 8 mm. The SLM is placed at the beam path after multiplexing. Turbulence phase screen is generated by a von Kármán power spectral density \cite{Lane92,Reis25}. Speckle is generated by a Gaussian-correlated random phase screen emulating a diffuser \cite{Goodman76,Duncan08,Gong19}. We adopt the strength parameter of turbulence as $D/r_0$, the beam diameter over the coherence diameter, and the speckle as the diffuser phase $\sigma_s$, with the grain size $\rm g$ setting. A beam of diameter $D$ spans approximately $(D/r_0)^2$ statistically independent coherence cells. Speckle is treated as a scattering-induced noise rather than as a wavefront aberration. The random seed is fixed within each sweep, so that varying $D/r_0$ and $\sigma_s$ changes only the amplitude of the screen while its spatial pattern is preserved. Varying $\rm g$ necessarily changes the pattern itself, since the grain size defines it. The affected demultiplexed amplitude is shown in Fig. 4. The examples of turbulence and speckle case are in inset. In (a) inset, turbulence parameter is $D/r_0=10$, and in (b) inset, speckle parameters are $\sigma_s=1$ rad, $\rm g=100\ \mu$m. The amplitude degradation in this section is measured from the CMOS camera images rather than with the photoreceiver array. Each image is processed as in Fig. 2, the azimuthal profile on the detection ring is demultiplexed and the harmonic amplitudes are read. At $D/r_0=10$ the share of the harmonic power in the odd orders, which the parity rejects, rises from 4.4 \% to 15 \%, while the spurious even orders above four rise only from 0.5 \% to 1.5 \%, and the two channel harmonics retain 83 \% of the harmonic power. Under speckle at $\sigma_s=1$ rad and $\rm g=100\ \mu$m the odd share stays at 4.5 \% and the spurious even share reaches 1.7 \%. 

We chose this over a direct BER measurement through the SLM because a reflective SLM inserts a fixed loss of its own, from its diffraction efficiency and fill factor, that is absent from the measurements of Fig. 3, so a BER measured through the SLM would not correspond one-to-one to that of Fig. 3. Instead, the SLM stays in the beam path throughout, the sweep starts from a flat phase screen that represents the undisturbed beam, and every amplitude is normalized to that reference. The fixed loss of the SLM cancels in the ratio, and what remains is the effect of the turbulence or speckle screen alone. We can see degradation of each channel relative to the undisturbed beam. These measurements quantify the dependence of the demultiplexed amplitude on turbulence and speckle strength, from which the SNR and BER trends of our system follow. 

We can calculate degraded scenario's SNR and BER to propose trade off. For example, at turbulence $D/r_0=10$, channel 1 is degraded 10.3 \% and channel 2 is degraded 9.0 \%. The azimuthally averaged intensity on the detection ring falls by 6.6 \% relative to the undisturbed beam. SNR falls by 0.95 dB on channel 1 and 0.82 dB on channel 2. At $\sigma_s=1$ rad and $\rm g=100\ \mu$m, channel 1 is degraded 9.0 \% and channel 2 is degraded 10.8 \%. The azimuthal averaged intensity on the detection ring falls by 6.7 \% relative to the undisturbed beam. SNR falls by 0.82 dB on channel 1 and  1.00 dB on channel 2, independently of the data rate since the degradation scales the amplitude and not the noise. Propagating these amplitudes through the unperturbed measured BER by scaling each channel's $Q$ by its amplitude ratio, recomputing the BER, and interpolating the FEC crossing in log BER between the measured points, moves the FEC limited rate from 5.2 to 5.05 Gb/s at $D/r_0 = 10$ and to 5.04 Gb/s at $\sigma_s=1$ rad and $\rm g=100\ \mu$m, a reduction of about 2.9 \% and 3.1 \%. This result is not a direct measured BER under turbulence. We can approximately predict trade-off point of data rate in disturbed scenario with these calculation process.

\section{Conclusion}

We have proposed and experimentally demonstrated a demultiplexing method for OAM light. Each channel carries a different OAM state as a VVB, and projection of the VVB onto a linear polarization state generates a petal pattern. The intensity profile of the petal pattern for topological charge $\ell$ carries the cyclic symmetry $C_{2\ell}$ and therefore a single azimuthal harmonic of order $p=2\ell$. With this property, each channel is demultiplexed by its own harmonic through a Fourier transform of the azimuthal intensity profile, without any additional optical component in the receive path.

A two-channel system is designed with $\ell=1$ and $\ell=2$ at 515 nm, multiplexed into a single beam and received on a circular photoreceiver array and CMOS camera. The two channels are recovered from the same set of samples. The system reaches 2.6 Gb/s per channel, 5.2 Gb/s in total, below the FEC threshold of $3.8\times10^{-3}$. The per-channel rate is 1.86 times the 1.4 GHz photoreceiver bandwidth, and the FEC-limited point is set by the bandwidth and responsivity of the photoreceiver rather than by crosstalk between the channels. The crosstalk is below $-17$ dB on channel 1 and $-7.3$ dB on channel 2, and the latter can be removed after detection with one subtraction per bit. Direct detection OAM communication without MIMO equalization have reported aggregate rates of 3 Gb/s over two modes through 2.96 m of water at 445 nm \cite{Baghdady16}, 2 Gb/s over two modes through aerosol and an air--water interface at 532 nm \cite{Song21}, and 2 Gb/s over two modes at 3.4 $\mu$m \cite{Zuo26}, each with a mode-converting element in the receive path and one photodetector per mode or one mode recovered at a time. The present system reaches 5.2 Gb/s over two modes with a photoreceiver ring and a Fourier transform. Its 2.6 Gb/s per channel through a 1.4 GHz photoreceiver, against 1 Gb/s per channel through receivers of 1.6 GHz \cite{Song21} and 1 GHz \cite{Zuo26}, shows that the Fourier-transform demultiplexer is not the element that limits the rate. Under emulated turbulence of $D/r_0=10$ and speckle of $\sigma_s=1$ rad and grain size $\rm g=100\ \mu$m, the demultiplexed amplitude of each channel falls by about 10 \%, corresponding to an SNR loss of about 1 dB independent of the data rate.

This method can be extended in several directions. The channel count is set by the azimuthal sampling. Channel $\ell$ occupies harmonic $p=2\ell$ and $N_{\rm d}$ number of sampling positions resolve harmonics up to $p=N_{\rm d}/2$, so the array supports $\ell\le N_{\rm d}/4$. 16 positions admit $\ell\le4$, and more photoreceivers admit higher $\ell$. Since the polarization remains free once the petal pattern is formed, polarization multiplexing doubles the channel count with a polarizing beam splitter and a second array. Both extensions keep the receiver as it is, a ring of photoreceivers and a Fourier transform. For future work, a photoreceiver of higher bandwidth and responsivity together with a signal amplifier would raise the FEC-limited rate directly. In this paper, disturbance is only generated by SLM holographic with short link distance. Real-time demultiplexing in hardware of detector array, BER measured under real atmospheric turbulence or under-water over a longer link, and extension to more than two OAM channels remain to be demonstrated.

\appendix
\renewcommand{\thesection}{Appendix \Alph{section}}

\section{Mutual coherence case}

When the channels are mutually coherent, their fields add before detection. The multiplexed state at coherence case can be represented as
\begin{equation}
\sum_\ell E_\ell(\phi)=\frac1{\sqrt{2\pi}}\sum_\ell \cos(\ell\phi)
\end{equation}
, where $\alpha=0$ and $E_\ell(\phi)\equiv\left(\langle H|\otimes\langle \phi| \right)|\Psi_{\text{VVB}}\rangle$. The $\alpha_\ell$ enter only as phases of each harmonic and not as its index, so we set $\alpha_\ell=0$ without loss of generality. The intensity profile and coherence terms are given by
\begin{equation}
\left| \sum_\ell E_\ell(\phi)\right|^2=\frac{1}{4\pi}\left\{N+\sum_{\ell\in \mathcal N}\cos\big(2\ell\phi\big)+2\sum_{\ell<\ell'} \underbrace{\big[\cos((\ell'-\ell)\phi)+\cos\big((\ell'+\ell)\phi\big)\big]}_{\text{coherence terms}} \right\}
\end{equation}
, where $\mathcal N$ is the set of multiplexed modes, and $N=|\mathcal N|$.

\section{Crosstalk definition for two channels}

The demultiplexed amplitude of channel $i$ is a linear combination of the bits sent on both channels,
\begin{equation}
s_i=\sum_j M_{ij}b_j+c_i
\end{equation}
, where $b_j\in\{0,1\}$ is the bit sent on channel $j$, $M_{ij}$ is the response of channel $i$ to channel $j$, and $c_i$ is a constant. The elements of $M$ are obtained by least-squares regression of $s_i$ on the transmitted bits. For the two channels,
\begin{equation}
M=\begin{pmatrix}M_{11}&M_{12}\\M_{21}&M_{22}\end{pmatrix}
\end{equation}
, and the crosstalk into channel $i$ is
\begin{equation}
\mathrm{XT}_i=10\log_{10}\left|\frac{M_{ij}}{M_{ii}}\right|,\qquad (j\neq i)
\end{equation}
Since the photoreceiver output is linear in optical power, $M_{ij}/M_{ii}$ is a ratio of optical powers and $\mathrm{XT}_i$ is the optical crosstalk. Channel 1 is decided from $s_1=M_{11}b_1+M_{12}b_2+c_1$ and channel 2 from $s_2=M_{21}b_1+M_{22}b_2+c_2$.

\section*{}

\noindent\textbf{Funding.} This work was supported by the Ministry of Science and ICT (MSIT), Korea, through Information Technology Research Center (ITRC) Program (IITP-2026-RS-2023-00259061); Institute for Information and Communications Technology Planning and Evaluation (IITP).
This research was supported by Basic Science Research Program through the
National Research Foundation of Korea(NRF) funded by the Ministry of Education(RS-2026-25573747).

\noindent\textbf{Disclosures.} The authors declare no conflicts of interest.

\noindent\textbf{Data availability.} Data underlying the results presented in this paper are not publicly available at this time but may be obtained from the authors upon reasonable request.

\end{document}